\documentclass[amsmath,amssymb,aps,prab,reprint,groupedaddress]{revtex4-2}
\usepackage{graphicx}
\usepackage{dcolumn}
\usepackage{bm}
\usepackage{subfig}
\begin{document}


\title{Partial Meissner effect measurement by a superconducting magnetic flux lens}

\author{A. Ivanov}
\altaffiliation[]{CERN, Geneva Switzerland}
\email{anton.ivanov@cern.ch}
\author{T. Koettig}
\affiliation{CERN, Geneva Switzerland} 
\author{A. Macpherson}
\affiliation{CERN, Geneva Switzerland} 
\date{\today}

\begin{abstract}
Magnetic flux trapping in the Meissner transition of superconducting radio frequency cavities can substantially increase dissipation which impacts cryogenic costs and necessitates expensive magnetic shielding. Recent findings point at the material preparation and the cool down dynamics at the vortex state as the main ways to counteract flux trapping. Most of the related measurements are performed on the cavity, with risk of results being impacted by the cool down specifics of each facility. We demonstrate an alternative experiment in which flux expulsion is quantified from sheet material, with samples conduction cooled in a stand-alone instrument designed to collimate expelled flux at controlled cool down conditions. A series of tests reliably and reproducibly relate the amount of expelled flux to the cooling rate, the velocity of the superconducting front and the spatial temperature gradient. The results are in line with tests performed on bare cavities and show that expulsion of trapped flux improves with the spatial temperature gradient, independent on the cooling rate. Our concept offers means to control the dynamics of the Meissner effect and can be used for material qualification prior to cavity fabrication.
\end{abstract}


\maketitle

\section{\label{sec:intro}Introduction}
The ability of a superconductor to expel magnetic flux when
cooled to the Meissner state has direct implications in the performance of superconducting radio-frequency (SRF) cavities. During the superconducting transition, defects in the cavity material can pin the externally present ambient magnetic field $B_0$ as quantized vortices, leading to magnetic flux being trapped in the superconducting material. Under the RF field in the cavity, this trapped flux can cause localized heating which, on the macroscopic scale, contributes to the residual surface resistance $R_0$. For applications requiring continuous or long-pulse accelerating gradients, efficient expulsion of trapped flux is therefore crucial to optimize cavity performance and minimize cryogenic losses \cite{Padamsee_2017}.

To mitigate $R_0$ due to trapped flux, a seemingly straight-forward approach is to shield the cavity from $B_0$. In practice, the high cost and various difficulties in engineering and magnetic hygiene limit achievable shielding performance \cite{Wu_2015}. Additionally, magnetic field of thermoelectric origin generated at bi-metal junctions may act as an internal source of trapped flux, both for bulk niobium \cite{Knobloch_2019,Eichhorn_2016} and niobium coated cavities \cite{Miyazaki_2019}. Thus, mitigation of $R_{\textrm{0}}$ cannot be addressed solely by magnetic shielding, and one has to optimise the intrinsic expulsion ability of the material.

It is well known that material preparation can affect expulsion by introducing bulk features such as impurities, dislocations or grain boundaries acting as pinning centers where flux vortices are trapped \cite{Antoine_2018}. Recrystallization by heat treatment of polycrystalline niobium has shown to significantly reduce flux trapping in samples \cite{Bahte_2009} as well as in cavities \cite{Posen_2016}, suggesting improvement in expulsion correlates with grain growth. However, single crystal niobium was found to have incomplete expulsion that could also be improved by heat treatment \cite{Aull_Knobloch_2012}, implying that grain boundaries are not the sole pinning source. Further, while a  heat treatment of $T \gtrapprox 1400^{\circ}$C allows nearly full expulsion in samples \cite{Laxdal_2018,Bahte_2009}, it also compromises the yield strength of niobium. At present, even with typical heat treatments of 800 to 1000$^{\circ}$C, SRF cavities retain the potential of non-negligible flux trapping.

To complicate things further, it has been experimentally recognized that cool down dynamics near the transition temperature $T_{\textrm{c}}$ affects the cavity performance \cite{Kugeler_2009}. Later studies have correlated flux expulsion to the temperature difference between the warm/cold ends of the cavity, with the spatial temperature gradient $\nabla T$ and the cooling rate as the key parameters considered. Surface resistance was observed to improve for slow cooling rates and low $\nabla T$ in dressed cavities in horizontal setups \cite{Vogt_2013,Valles_2013} with the trapped flux attributed to thermoelectric effects generated between the cavity and the liquid helium tank \cite{Eichhorn_2016,Knobloch_2019}. Similar behavior seen in performance tests of niobium coated cavity also points at thermoelectric effects \cite{Venturini_2015}. On the contrary, for bare cavities in vertical setups, fast cooling and high $\nabla T$ were found to increase expelled flux \cite{Romanenko_2014,Romanenko_Sergatskov2014,Kubo_2016}, suggesting this behavior is intrinsic to the material itself, as in this case thermoelectric effects could be excluded. Two models were proposed to explain the underlying mechanism: $\nabla T$ promotes expulsion by acting as a depinning force at the superconducting/normal conducting (SC/NC) front  \cite{Checchin_2016}, or alternatively by reducing the vortex region and in turn the probability of a trapping event  \cite{Kubo_2016} -- both suggesting the fast cool down approach to improve flux expulsion for bare cavities in vertical test cryostats \cite{Posen_2016,Dhakal_2020}.  

\subsection{Measurement of flux expulsion}
While methods based on magneto-optics \cite{Vinnikov_1982} and polarized neutron radiography \cite{Kugeler_2017} allow  spatially resolved imaging of trapped flux, the majority of expulsion measurements in the SRF community are somewhat based on the original Ochsenfeld experiment: In the Meissner state screening currents induce a magnetization that tries to cancel $B_0$ inside the material. Due to flux conservation,  a geometry-dependent change of the flux density distribution around the superconductor is observed and is commonly referred to as expulsion ratio (ER). In the ideal case of a complete Meissner expulsion, the superconductor can be approximated to a perfect diamagnet with zero flux density inside and $B_{sc}'$ at a given point outside the material. The latter can be evaluated numerically or, alternatively, measured by zero-field cooling followed by $B_0$ ramping \cite{Benvenuti_1999}. However, for a realistic cool down of a cavity material such as niobium (a type-II superconductor) with $B_0 \neq 0$, the flux condensates in the form of vortices arising in the mixed state localized within few hundred microns around the SC/NC interface \cite{Kubo_2016}. As the transition front propagates, this mixed state region is swept throughout the material with flux vortices either being transported with the front or pinned locally.  At the Meissner state, pinned vortices contribute to a localized cancellation of the screening currents, thereby modifying the macroscopic flux density near the superconductor i.e. $B_{sc} \neq B_{sc}'$. The latter offers an observable of trapped flux that can be resolved via magnetometry.

Expulsion experiments based on magnetometry are typically performed during a bare cavity test in a vertical cryostat where active coils are used to control $B_0$ and the mass flow rate of the helium bath is somewhat used to control the cool down dynamics. For elliptical cavities with $B_0$ along the main cavity axis and $B_{sc}'/ B_{0}$ measured at the equator, values of $\approx{1.8}$ indicate ideal expulsion (or 100\% expulsion efficiency), however interpretation can be more involved in the complicated low-beta \cite{Miyazaki_2020} or crab \cite{Hernandez_2015} geometries. While such experiments are valuable for the SRF community to correlate the amount of trapped flux to the RF performance, well defined cooling conditions and magnetic field environment are not trivial, with results impacted by the cool down protocol applied in each facility. 

In addition to cavity tests, dedicated setups were also used to perform magnetometry on flat samples \cite{Aull_Knobloch_2012} and a bar of cavity-grade niobium \cite{Vogt_2013}. Results indicated that flux trapping was related to cooling rate, but contrary to the majority of bare cavity results, slow cool down was found to improve expulsion. However, as the focus of these earlier studies was on thermoelectric effects, $\nabla T$ could not be unambiguously related to the intrinsic expulsion performance of the material.  
   
\subsection{Goal of the study}
In this work we aim at a dedicated experiment to systematically relate flux expulsion to the history of material preparation and the cool down dynamics near $T_c$. A stand-alone magnetometry measurement performed on a compact sample allows rapid quantitative assessment of raw material prior to cavity fabrication. The current consensus between measurements and theoretical models suggests such an experiment should unambiguously relate expulsion to $\nabla T$ decoupled from other the dynamic quantities and thermoelectric effects.
A compact setup would in principle facilitate the control of the cryo-magnetic environment, however it also implies a trade-off in the amount of flux to be expelled in the transition, and increased sensitivity to edge effects on the field distribution near the superconductor. These issues are resolved by developing an instrument intended to collimate flux by cooling on a closed thermal topology. This allowed direct observation of the SC front evolution with resolution similar to  experiments performed on cavities. The idea is explained in section \ref{concept}. Section \ref{Proof of concept} is a proof of concept description of the prototyped instrument and the setup followed by results and discussion from benchmark experiments. Finally, in section \ref{Summary} we summarize and point to potential applications.
\begin{figure*}
	\includegraphics[scale=0.4]{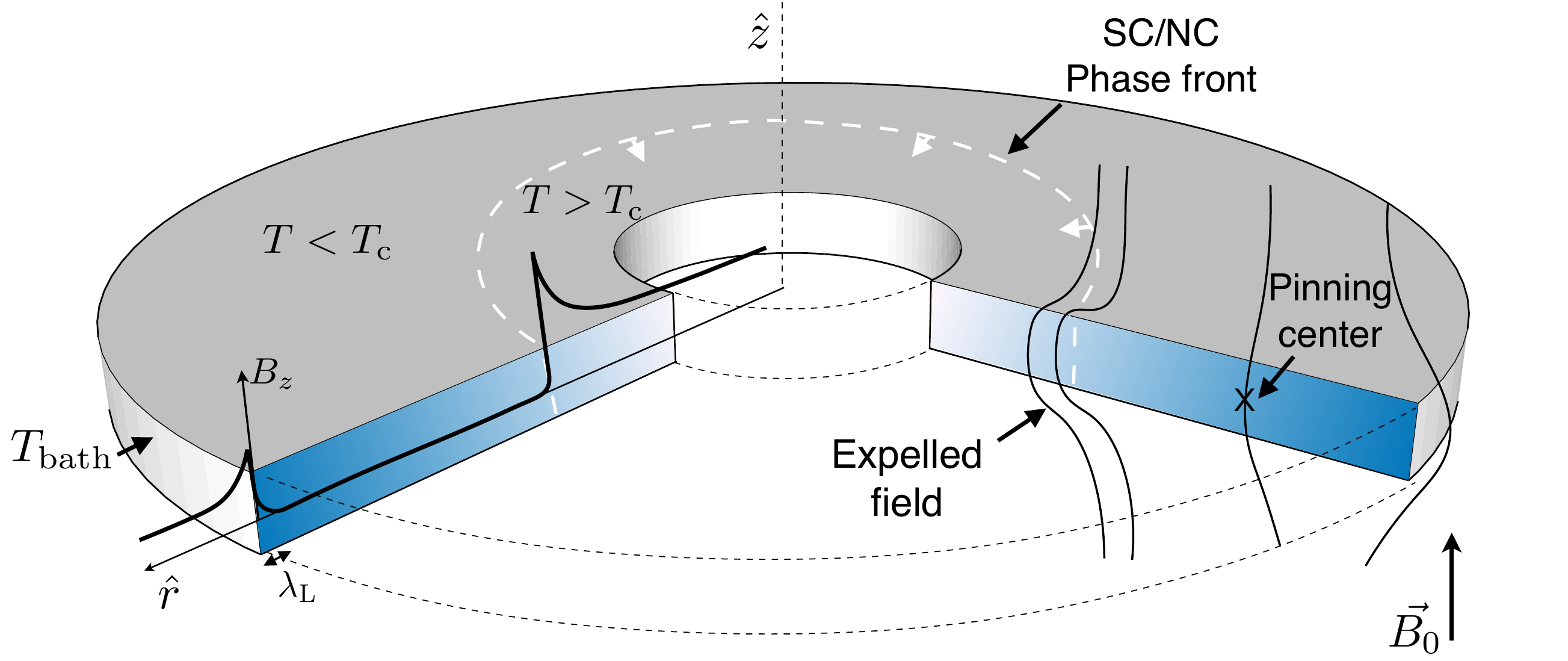}
	\caption{\label{disc_1_details} Axially-symmetric geometry proposed to measure flux expulsion parametrized as: $d$ -- inner diameter; $D$ -- outer diameter; $w$ -- thickness such that $\lambda_{\textrm L} << w << D$, where $\lambda_L$ is London penetration depth. The radial profile and the color gradient shown at the cross section illustrate the magnetic flux density and temperature distributions. 
	}
\end{figure*}    
 
\section{CONCEPT}
\label{concept} 
The geometry adopted in our experiment is shown in figure \ref{disc_1_details}. The sample under test is a circular disc of sheet niobium  perforated with a circular aperture at the center. The ambient magnetic field $B_0<B_c$ is
applied along its symmetry axis ($z$-axis in figure \ref{disc_1_details}) such that in the NC state the field lines penetrate through the material perpendicular to its main surface. The outer edge of the sample is preferentially cooled through a thermal contact to a cold mass at the disc's outer surface. As the temperature  of the cold mass is lowered, the periphery of the disc is the first to be cooled below the critical temperature $T < T_c$. In turn, a thin closed-loop ring of a SC element is formed that topologically traps all the flux permeating the remaining NC inner part of the disc. As cool down proceeds, this SC ring expands radially inward collimating the flux. Once the entire disc is transitioned to the SC state, flux has been expelled from the sample volume and is \textit{focused} at the on-axis aperture, and for this reason the setup is referred to as a flux expulsion lens.

Given the above assumptions, the change in flux density integrated over the aperture area can be measured by a coil (known as a search coil sensor \cite{Buzio_2011}) for which the induced voltage offers a direct observable of the amount of flux expelled over the sample. In this case expulsion efficiency is simply deduced from: $ \Delta \Phi_\textrm{c}(\nabla T)  / \Phi_0$ where $\Delta \Phi_\textrm{c}$ is the amount of flux change detected by the coil and $\Phi_0$ is the flux across the sample area in the NC state, which is known assuming $B_0$ is sufficiently homogeneous. 

\subsection{Magnetostatic simulations}
\label{Magnetic_response}
\begin{figure}
	\includegraphics[scale=0.54]{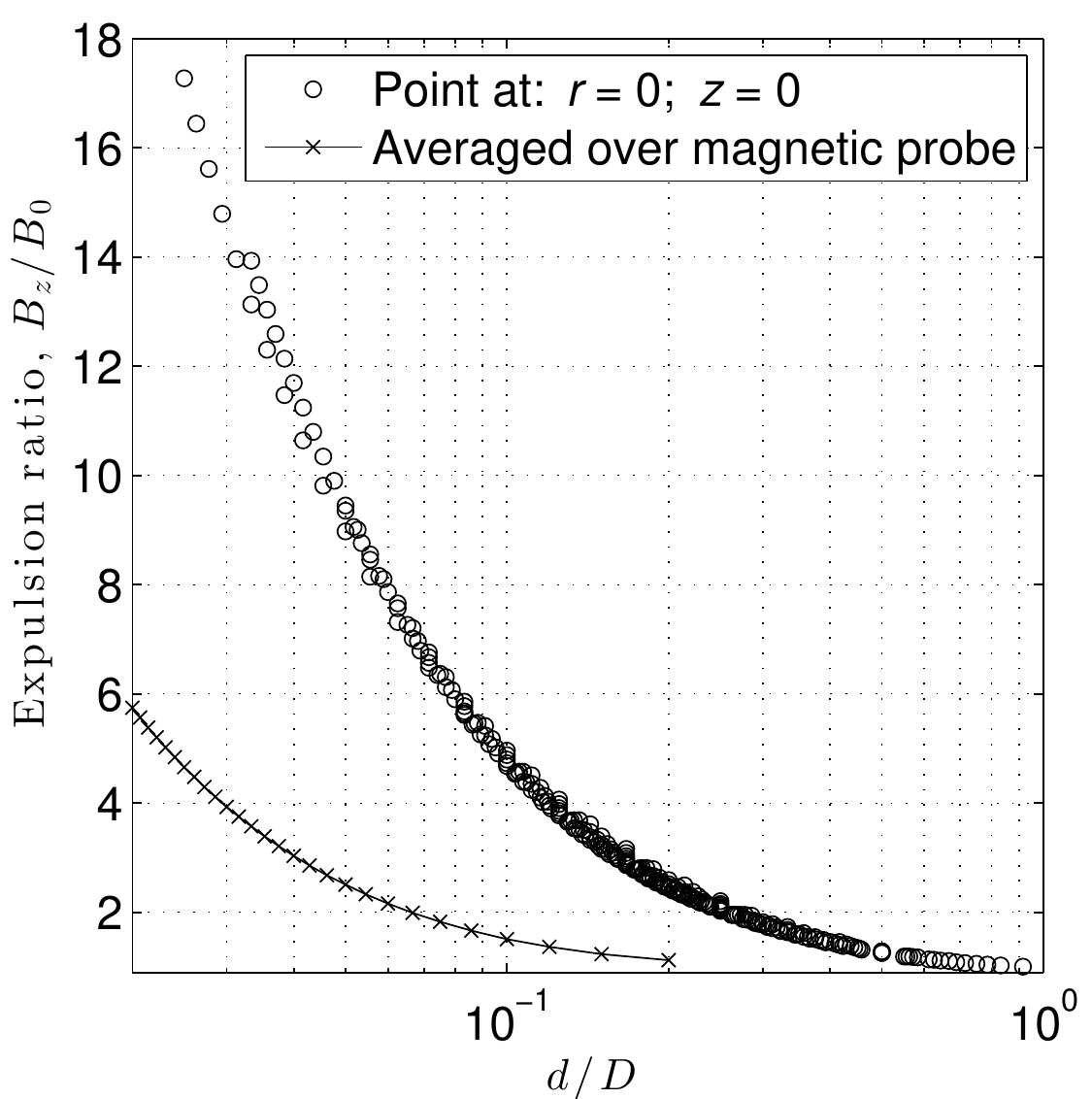}
	\caption{\label{exp_factor_2d} Expulsion ratio as function of the relative aperture diameter: ({upper}) obtained at the aperture center; ({lower}) obtained for a disc aperture of 0.6 cm from the mean flux density evaluated over the probe volume. }
\end{figure}

For the purpose of this study no search coil was implemented, instead we measure magnetic flux density using a commercially available flux-gate probe, for which the ER is $B_{\textrm{sc}}/B_0$ measured at the aperture. To quantify the maximum ER   attributed to material that would exhibit ideal Meissner effect,  magneto-static FEM simulations were performed using COMSOL\textsuperscript{\textregistered}, with the sample treated as   $\mu_r = 0$ material. Based on these results, with examples shown in figure \ref{Field_distribution}, we can make the following observations: 
\begin{figure}
	\includegraphics[scale=0.26]{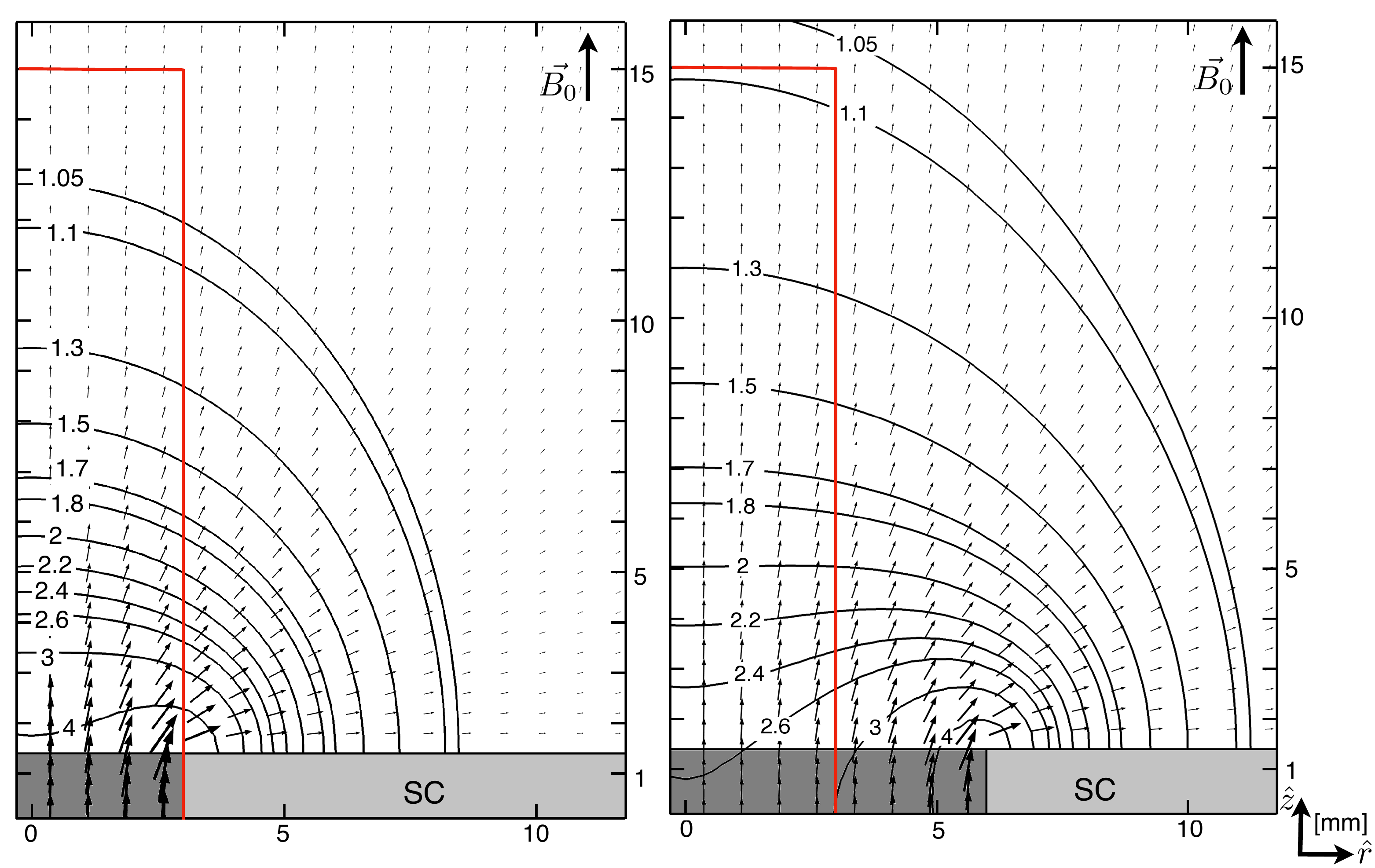}
	\caption{\label{Field_distribution} 
		Simulated field distribution (arrows) shown near the main axis of a SC disc, with $D = 9$ cm and $d=0.6$ cm (left), and $d=1.2$ cm (right); Contours are levels of $B_{z}/B_0$; Red lines represent the flux-gate probe position.}
\end{figure}
\begin{enumerate}
	\item In the SC state, at the vicinity of the aperture, the field is oriented predominantly along the disc's main axis. A single axis flux-gate magnetometer centered at the aperture is hence sufficient to measure the density increase attributed to flux expulsion.
	\item As expected from flux conservation, reducing the relative size of the aperture $d/D$ increases flux density. To quantify this effect we can use the result in figure \ref{exp_factor_2d} ({upper curve}) shown for the flux density found at the center of the aperture.  If we consider the Bartington cryogenic Mag-01H fluxgate (depicted by the red contour in figure \ref{Field_distribution}) it is evident that the field varies substantially along the probe as its length is nearly ten times greater, compared to the thickness $w$. The field density measured by the probe is then averaged over the probe volume which suggests that the ER will be effectively reduced. In an attempt to quantify this effect we consider the mean value of the on-axis flux density component evaluated over multiple points within the volume of the probe: the {bottom curve} shown in figure \ref{exp_factor_2d}. 
	\item With an aperture set by the diameter of the flux-gate probe ($d=6$ mm), for the outer diameter $D$ in the range of 3 cm to 9 cm a complete flux expulsion would give ER between 1.1 and 2 -- somewhat comparable to expulsion experiments performed on various elliptical type cavities. 
\end{enumerate}
While these very simplified simulations are somewhat useful to provide us with design guidelines, as they do not account for the complexities of the SC transition and flux trapping it is not straight-forward to relate the absolute amount of expelled flux to the amount of field enhancement measured by the probe. 

\subsection{Thermal control}
\label{thermal_control} 
\begin{figure}
	\includegraphics[scale=0.54]{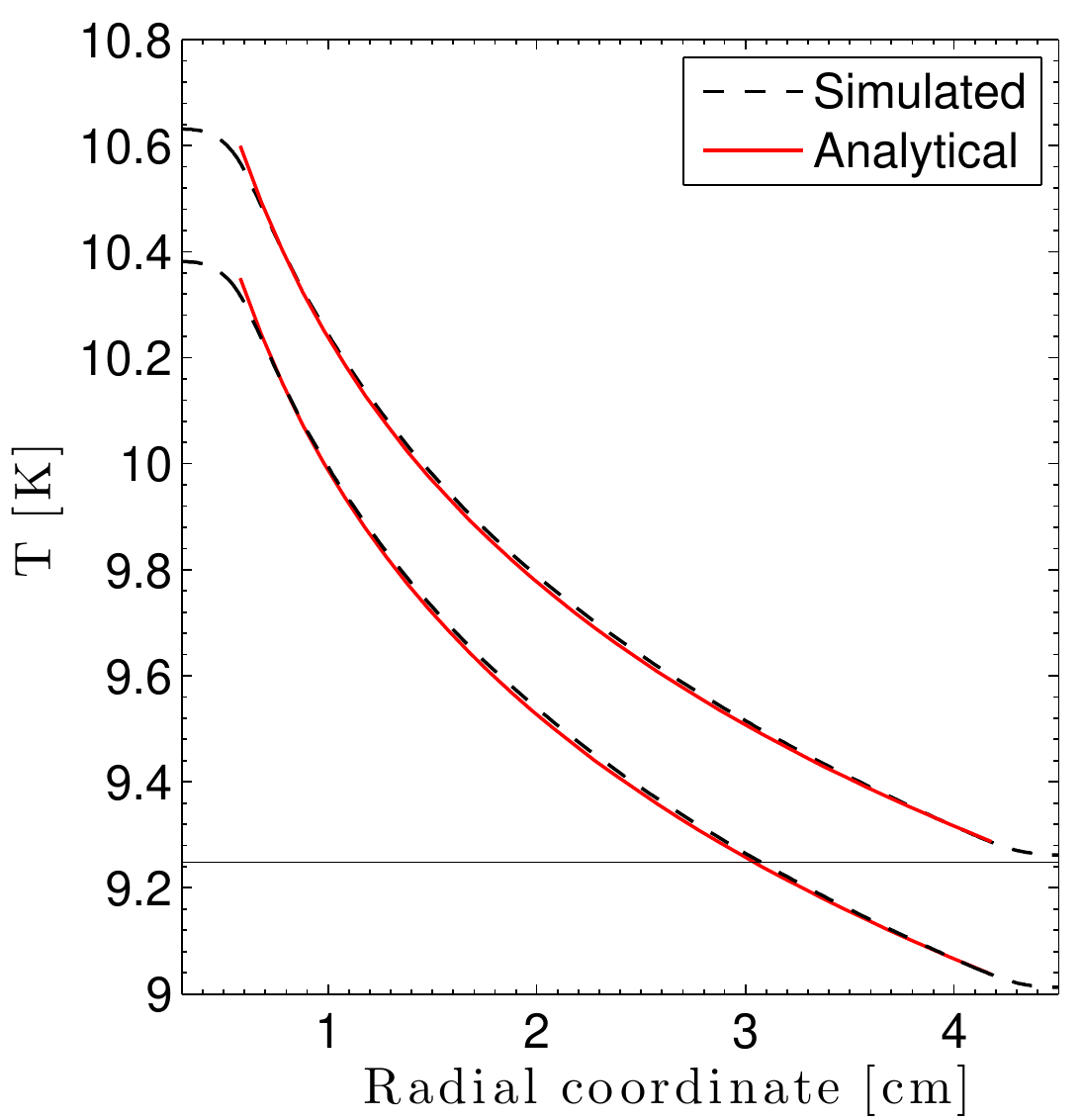}
	\caption{\label{T_profiles} Steady radial temperature distribution calculated for a sample with: $d = 0.6$ cm and $D = 9 $ cm. Simulated result was obtained via COMSOL\textsuperscript{\textregistered} where the thermal conductivity of niobium $k(T)$ is taken \cite{Julia_Thesis} and perfect thermal interfaces are assumed. The analytical result is the steady solution of the heat equation for annular geometry with constant thermal conductivity and $T_d$, $T_D$  applied at the inner and outer circular surface.}
\end{figure}

To drive the SC transition radially, a sufficiently uniform temperature control is required so that the closed thermal topology is preserved during the cool down. Moreover, as we aim at characterizing ER as a function of $\nabla T$, a reproducible thermal conditions are needed as to systematically reach a range of $\nabla T$. 
To meet these requirements, two thermally attached ring-shaped baths with temperatures $T_D < T_d$ are located concentrically at the outer edge and around the aperture. In practice, this can be realized by copper tubes pressed against the flat surface of the sample as to ensure good thermal contact. In vacuum, the rest of the sample area can be considered adiabatic, with radial temperature distribution dictated solely by both boundary temperatures $T_D$ and $T_d$, as shown in figure \ref{T_profiles}. From the figure we see that, apart from a small geometrical effect coming from the finite thickness of the thermal baths, the stationary temperature distribution is well described by the analytic solution. Moreover, since near $T_c$, the thermal capacity of niobium is relatively low, transient effects can be neglected. This implies that the material responds immediately to change in the boundary temperatures, which permits the use of the steady solution to describe the evolution of the SC front in our conditions.

In short, controlling the difference between the initial temperatures at the two boundaries and their rate of change should give sufficient control over the spatial thermal gradient during the evolution of the transition front. In reality however, the thermodynamic behavior will be altered as the boundary temperatures are not independent but are modified by the properties of the thermal contacts. 
\section{Proof of concept} 
\label{Proof of concept}
A closed cycle cryocooler is used to implement the flux lens concept in practice, with the sample conduction cooled in vacuum. The temperature can be controlled and measured precisely so the desired cool down procedure is easier to achieve. As the sample can be thermally cycled within minutes, a large set of phase transitions can be obtained during a single test and without the added cost and time associated with liquid helium supply. 

\subsection{Flux lens design and prototype deployment}
\begin{figure}
	\includegraphics[scale=0.11]{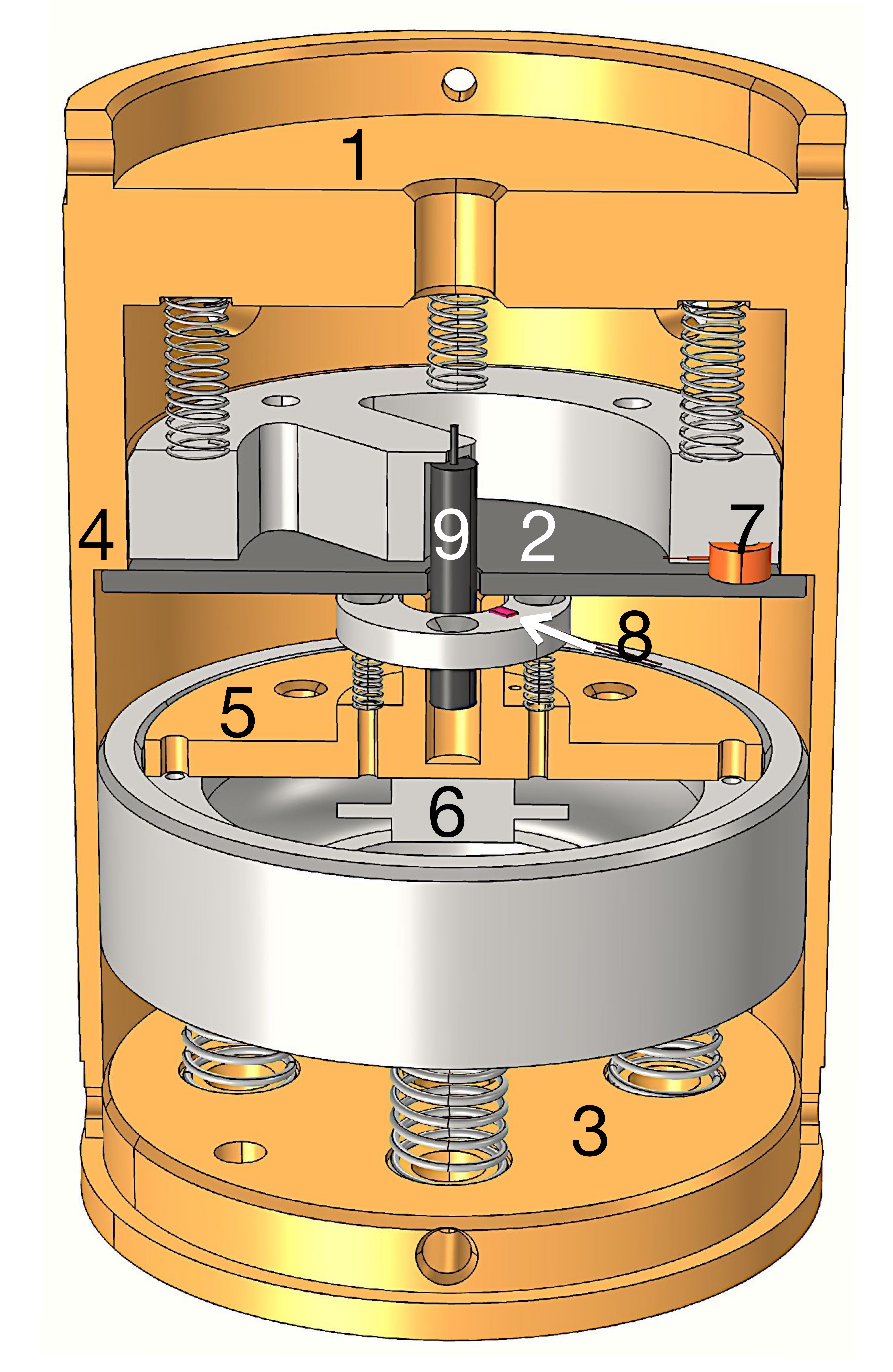}
	\caption{\label{flux_lens_design} Schematic of the developed flux lens design: 1 -- cryocooler interface; 2 -- Nb sample; 3 -- detachable lid; 4 -- cold edge; 5 -- sample heater; 6 -- heating resistor; 7 -- temperature sensor CERNOX\textsuperscript{\textregistered} (CU package);  8 -- temperature sensor CERNOX\textsuperscript{\textregistered} (SD package); 9 -- magnetic field probe.  }
\end{figure}

The design  of the flux lens shown in figure \ref{flux_lens_design} was developed to be compatible to a readily available cryocooler. The main body is a copper cylinder that serves as a cold mass and is firmly attached to the cold interface of the cryocooler. The sample is inserted from the opened end of the cylinder and rests on a step machined at it's inner wall. This provides a thermal path through which the outer edge of the disc is cooled (referred as cold edge hereafter). The heater used to warm the sample is effectively a copper pipe with a 10 W resistor mounted at one end. The radius of the pipe is chosen such that the thermal contact is formed locally around the aperture. Thereby, the heat from the resistor is transferred to the central part of the disc and a NC spot can be formed. Modulating the current amplitude through the resistor allows to sweep the size of the spot and in turn control the expulsion.  As the heater is spring-loaded the sample is kept firmly sandwiched between the cold edge and the heating pipe so that a good thermal contact can be maintained throughout operation. For thermometry of the sample, thermal sensors are mounted in  two spring-loaded holders at the upper and lower surfaces of the sample near the cold edge and the aperture, again with direct thermal contact assured. In this way all instrumentation is kept fixed which in turn helps keeping conditions unchanged across tests and makes sample replacement much easier. 
\begin{figure}
	\includegraphics[scale=0.31]{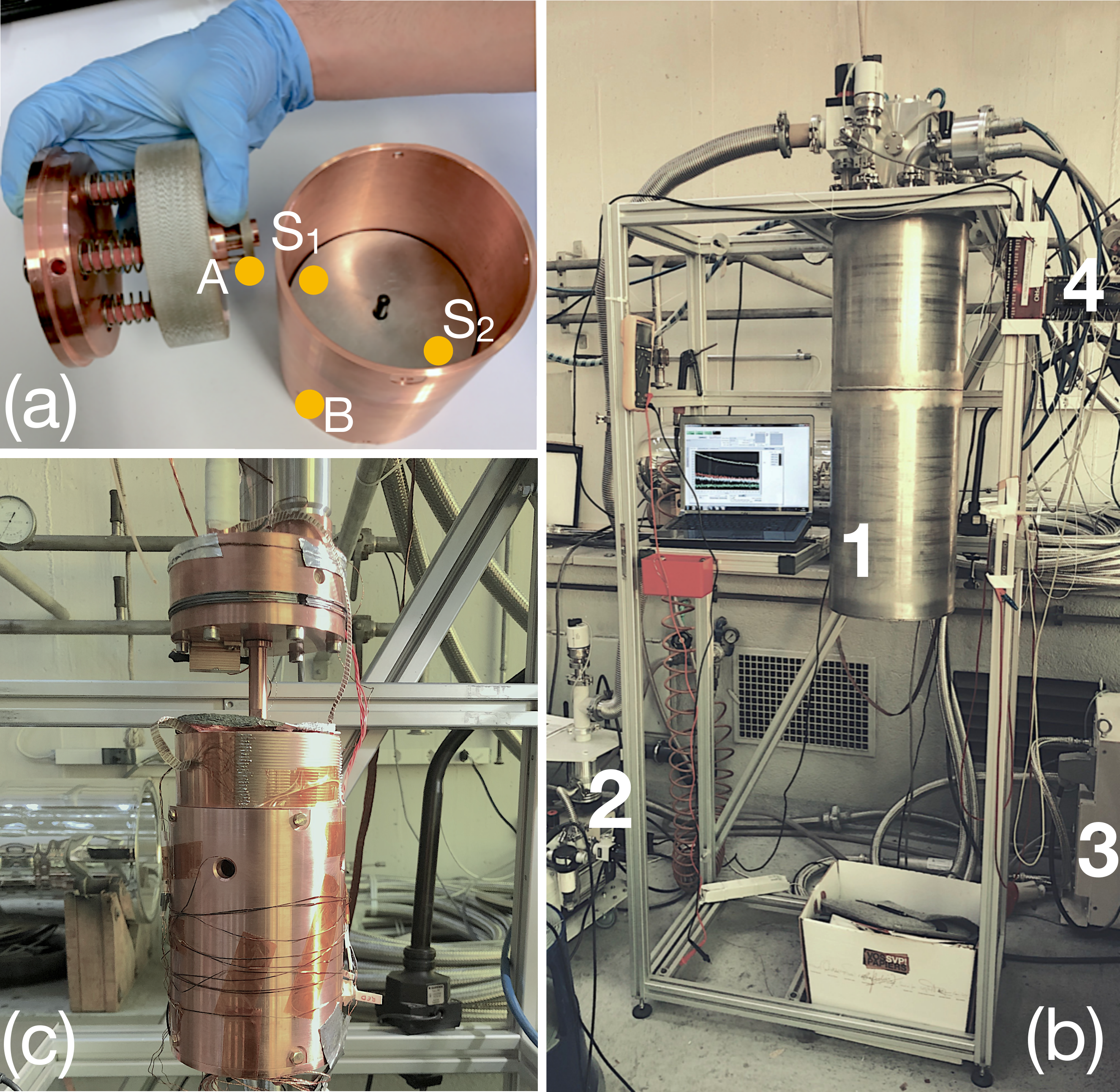}
	\caption{\label{setup_all} (a) Prototype shown along with the positions of the temperature sensors. (b) Setup used to perform the flux lens experiments: 1 -- vacuum vessel; 2 -- vacuum pump Pfeiffer PT50; 3 -- cryocooler Sumitomo SHB; 4 -- data acquisition hardware -- Labjack\textsuperscript{\textregistered} T7. (c) Flux lens mounted to the cold end of the cryocooler. }
\end{figure}

Based on the above-described design, the prototype shown in figure \ref{setup_all}a was developed. For sample dimensions matched to  the available cryocooler, the  largest possible sample diameter was  $D = 9$ cm, with $d =  0.6$ cm set at the smallest aperture compatible with the magnetic probe. Given that presently available  cryocoolers are limited in cooling power ($\approx 1$ W at 4 K in our case), the range of  spatial thermal  gradients is restricted and the choice of dimensions is intended to maximize the resolution where a complete flux expulsion, as roughly estimated in section \ref{Magnetic_response}, is  ${B_{sc}'}/{B_{0}} \approx 2$.
\subsection{Instrumentation and setup}
To measure magnetic flux density at the aperture a single-axis Bartington\textsuperscript{\textregistered} cryogenic Mag-01H magnetic probe is used and is aligned to the main axis of the sample. The temperature detection is resolved by a total of four CERNOX\textsuperscript{\textregistered} sensors placed  as indicated in figure \ref{setup_all}a: 
two CU-type ($\textrm S_1, \textrm S_2$) and a single smaller SD-type (B) are kept in contact with the surface of the sample (indium interface used) at the cold edge and near the aperture; another CU-type sensor (A) is attached to the flux lens body and is used to measure the cold mass temperature. All sensors are synchronously logged with a readout rate of $\approx 15$ Hz, which provides sufficient resolution for the transition dynamics.  To reduce thermal contractions during cool down and in turn degradation of the thermal contact, G10 fiberglass was used for all sensor holders. 

The setup was implemented at CERN's Central Cryogenic Laboratory and is shown in figure \ref{setup_all}b. At this stage, there is no magnetic shielding or active compensation coils available, thus all measurements were conducted at $B_0$,  effectively determined by earth's intrinsic field.
Given the setup orientation, the vertical component of the field is perpendicular to the sample's top and bottom surfaces and it was found to be $\approx 26$ $\mu$T as measured inside the cryocooler volume. The effect of the additional transverse component was simulated and it was found to not affect the expulsion ratio significantly as long as the probe is along the main axis of the disc. 

The possibility that a thermoelectric current is generated due to the bi-metal structure formed by the niobium disc and the flux lens body has been considered in the design, and in order to break the current path, all compression  springs are electrically decoupled so that the heater unit is kept floating. By symmetry, any thermoelectric potential that may form between the cold edge and the heat pipe is rendered as inconsequential as it cannot contribute to a net current. This was later confirmed throughout the multiple cool downs of the flux lens in which the magnetic field signature of such thermoelectric effects was typically around $0.4 \ \mu$T, hence significantly less than $B_0$. We therefore consider that in our setup thermoelectric currents are of no significance. 

The experimental procedure starts with installing the sample after which the flux lens is attached to the cryocooler interface. This is followed by pumping  (typically overnight) until the pressure inside the vacuum vessel is lowered to few $10^{-6}$ mbar. It then takes $\approx 6$ hours of cooling until the first expulsion is seen and few more hours to thermalize and reach the lowest temperature allowed by the crycooler ($\approx$ 3.9 K as measured at the cold mass). The sample measurement then proceeds  without interruption via an automatic procedure using heat pulses, giving systematic flux expulsion measurements at rate of $\approx$ 70 per day worth of measurement.
\begin{figure} 
	\centering
	\subfloat[Short heating pulse followed by a strong flux expulsion.] {\label{short_pulse_strong_expulsion}\includegraphics[width=0.95\columnwidth]{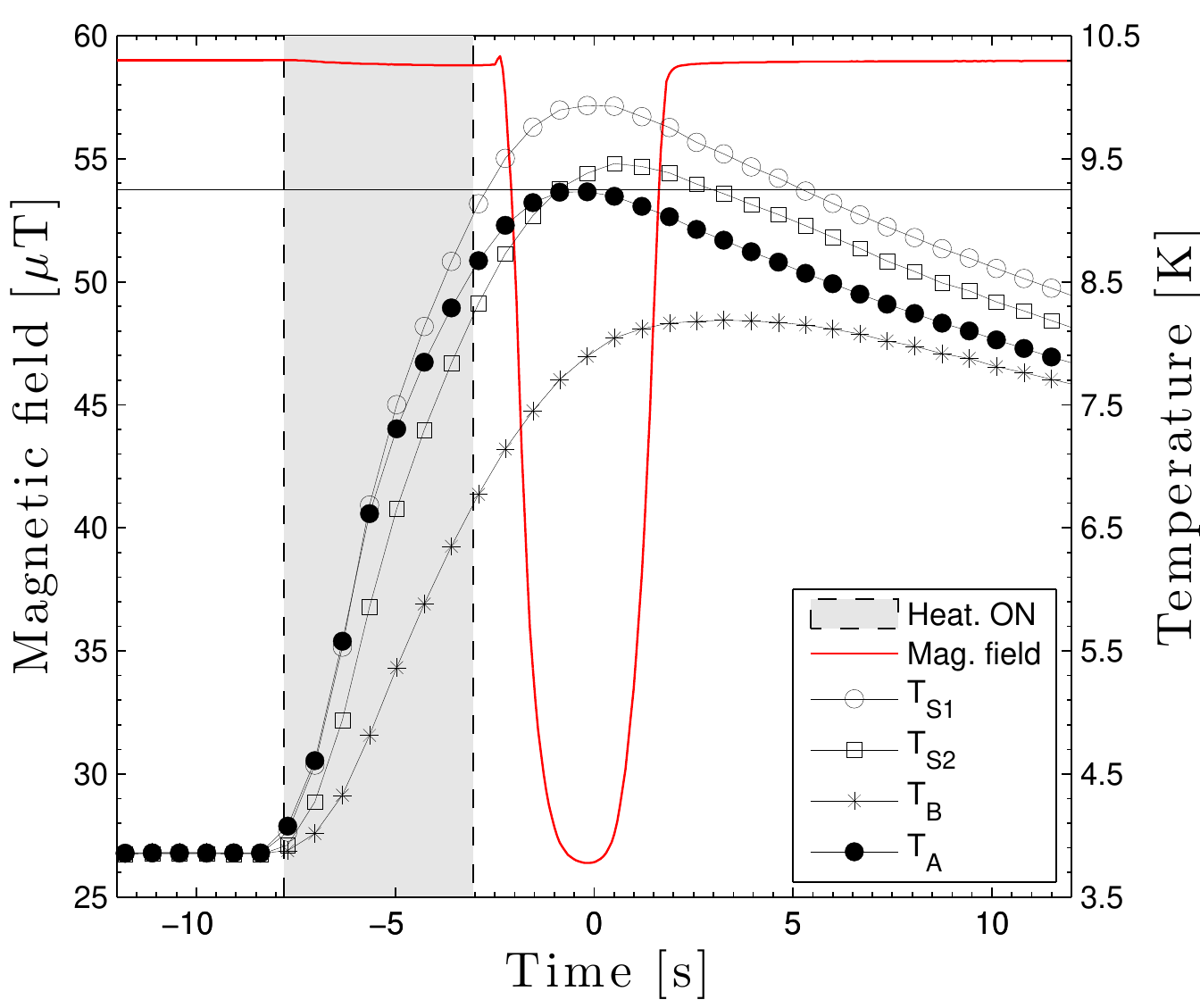} }\\ 
	\subfloat[Long heating pulse followed by a weak flux expulsion.] {\label{long_pulse_weak_expulsion}\includegraphics[width=0.95\columnwidth]{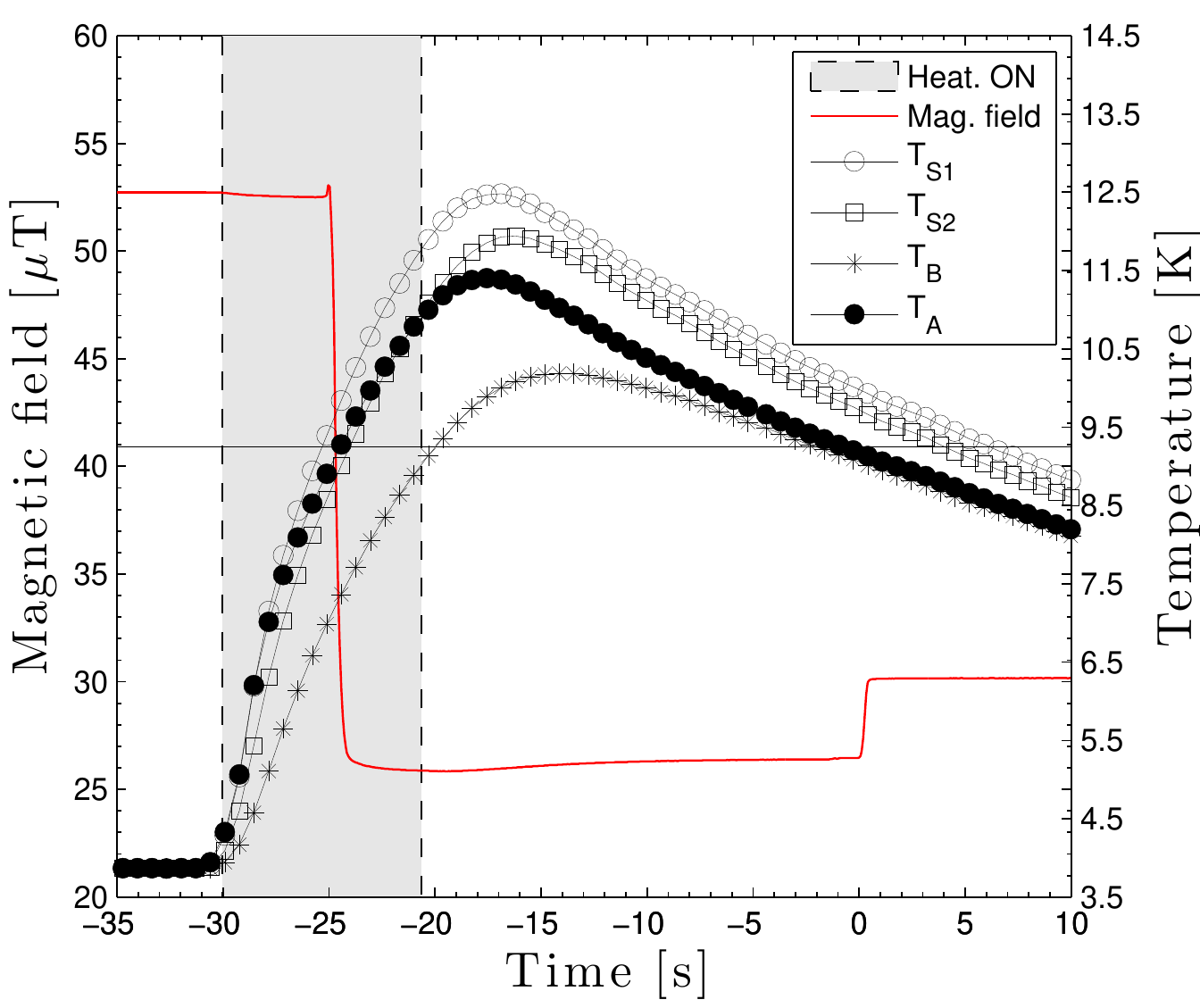} } 
	\caption[]
	{Magnetic field and temperature as measured during the course of a single flux expulsion experiment. Time axis is offset so that $t=0$ denotes the start of the expulsion.  The shaded area represents the duration of time for which the sample is heated. The offset among the temperature values measured at the sample surface indicates that some variation in the thermal contacts was present.} 
	\label{two_expulsions}
\end{figure}

To see how a single expulsion proceeds, figure \ref{short_pulse_strong_expulsion} can be used to illustrate: at $t = -10$ s the sample is SC and in thermal equilibrium with the cold mass at $ \approx 3.9$ K. A heating pulse of $ \approx 4$ W is then applied for $ \approx 4.5$ s, as indicated by the shaded area. In turn, the sample temperature rises and within seconds crosses $\textrm{T}_{\textrm c}$. This erases the magnetic memory from the preceding expulsion as shown by the decrease in  magnetic field amplitude to the ambient value of $\approx 26 \ \mu$T, indicating that the whole sample is NC. Once the heating pulse is turned off, the disk starts to thermalise with the cold mass, and the temperature drops quickly so that the sample re-transitions to the SC state within a few seconds, with the SC front advancing inwards driven by the cold mass cooling. In terms of magnetic field, it  starts increasing from $t = 0$ and then saturates about 1.8 seconds later at around 58 $\mu$T indicating that the whole volume of the sample is fully SC and flux expulsion completed. This pulse procedure can be repeated, but a systematic measurement requires that the cold bath temperature $T_{\textrm{A}}$ first drops again to its equilibrium value of $ \approx 3.9$ K.

The expulsions shown in figures \ref{short_pulse_strong_expulsion} and \ref{long_pulse_weak_expulsion} are performed according to the above-mentioned procedure under the same cryogenic and heat pulse conditions, with  only the length of the heating pulse is doubled in the second case. In the later, a significantly weaker expulsion is observed:  ER of $\approx$ 1.15 relative to $ \approx$ 2.2 in the former case. This implies that most of the flux was trapped in the material, suggesting that during the transition to the SC state, the sample was exposed to a lower temperature gradient. Longer heating lowers $\nabla T$ by bringing the sample closer to thermal equilibrium as more thermal energy arrives at the cold mass before the moment of expulsion. The heating pulse duration therefore offers a single parameter to perform multiple expulsions at variable $\nabla T$. It is however noted that the maximum achievable gradient in this case is limited by the minimum amount of thermal energy needed to warm the whole sample to $T>T_\textrm{c}$. For a fixed power, this sets the minimum heating pulse duration, which is determined experimentally by making sure the field decreases fully to $B_0$.  

To measure the temperature gradient we consider the thermal contact is uniform enough such that the disc is homogeneously cooled towards the center and the gradient has only a radial component: $\nabla T = \hat{r}{dT}/{dr}$. As cool down progresses the SC front (isothermal $T = T_\textrm{c}$ contour) propagates towards the aperture with the velocity also directed along the radius. This reduces the problem to one dimension and allows to utilize a previously proposed model \cite{Huang_Kubo_2016} by which the gradient at the position of the temperature sensor is obtained as:    

\begin{equation}
\label{gradT}
\frac{dT}{dr}=\frac{1}{\nu_{SC}}\frac{dT}{dt},
\end{equation}
where $\nu_{SC}$ is the speed of the SC front and $dT/dt$ is the cooling rate both considered when the SC front is at the position of the sensor. Because of the thermal topology, the rise of the magnetic probe signal offers a good observable for the time evolution of the SC front and can be used to obtain $\nu_{SC}$ by measuring the duration for expulsion.  To do so, we consider the time separation between the 10 \% and 90 \% levels up the rising edge of the probe signal during a given expulsion. The obtained values are found in the range of 0.25 s to 2.3 s, and correspond to the speed in the range of 18 cm/s to 2 cm/s, respectively. 
For the evaluation of the cooling rate the time signal from sensor A was found to be the most useful. Due to its smaller footprint it is characterized  by much lower thermal inertia and it was also easier to obtain consistent thermal contact with the disc surface. The values obtained for the cooling rate are found in the range of 0.2 K/s to 0.25 K/s. 

To test the flux lens a benchmark sample was prepared made of cavity-grade RRR = 300 niobium, cut via electrical discharge machining from a Nb sheet \footnote{material provided by Plansee Powertech AG}. A high expulsion efficiency is typically attributed to such un-worked niobium as it is in principle characterized by few pinning centers. 
For the proof-of-concept measurements both the  repeatability and reproduciblity  of expulsions in a range of $\nabla T$ have been considered. The former is defined as enough agreement between independent expulsions performed within a sample test. The latter is the agreement between results among independent tests: that is after the sample is warmed to room temperature, the flux lens disassembled and the sample remounted. 

\subsection{Results and discussion}
\label{Results and discussion}
\begin{figure}
	\includegraphics[scale=0.58]{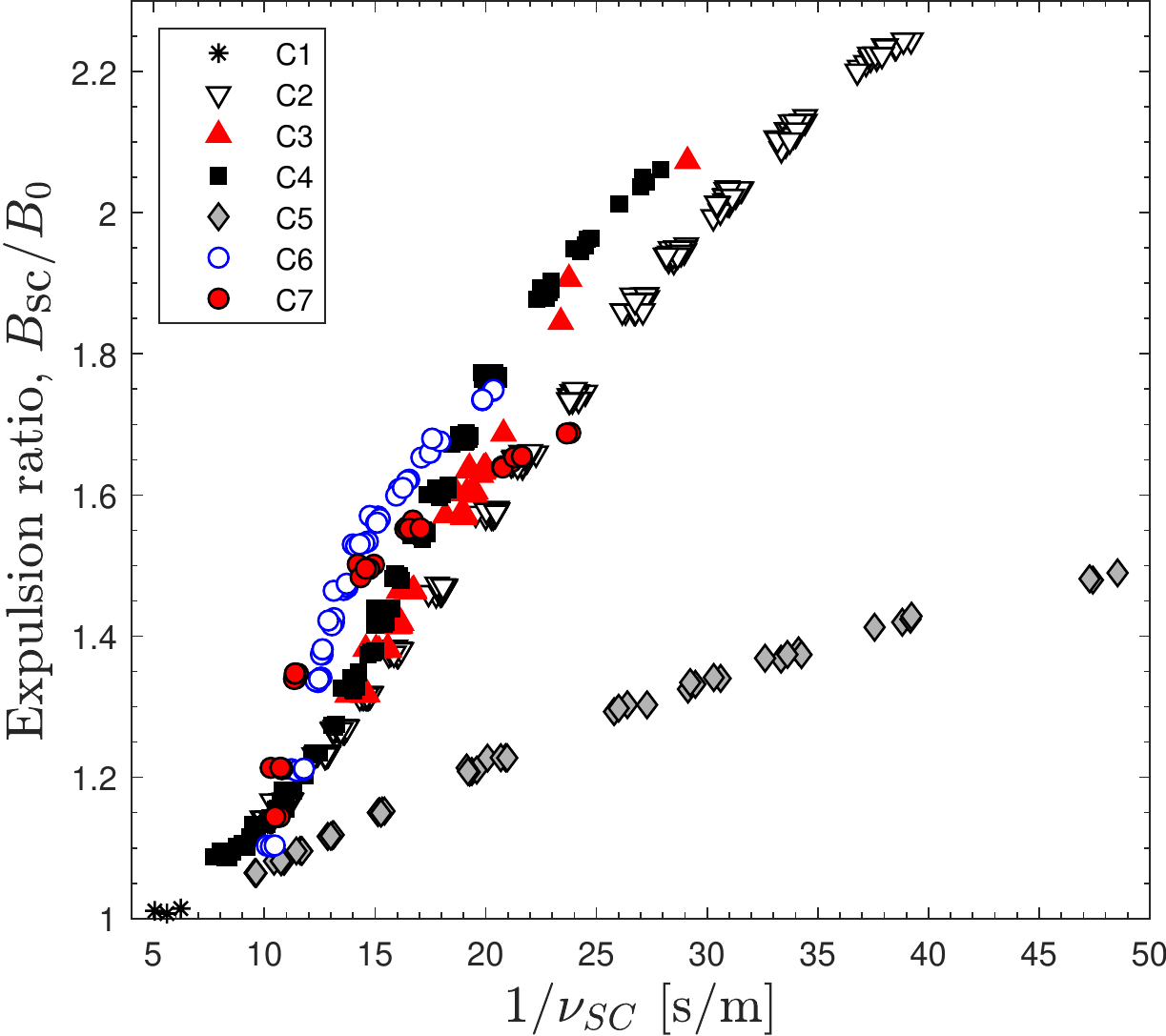}
	\caption{\label{EF_vs_front_speed} Expulsion ratio as a function of the inverse speed of the SC front.}
	\includegraphics[scale=0.58]{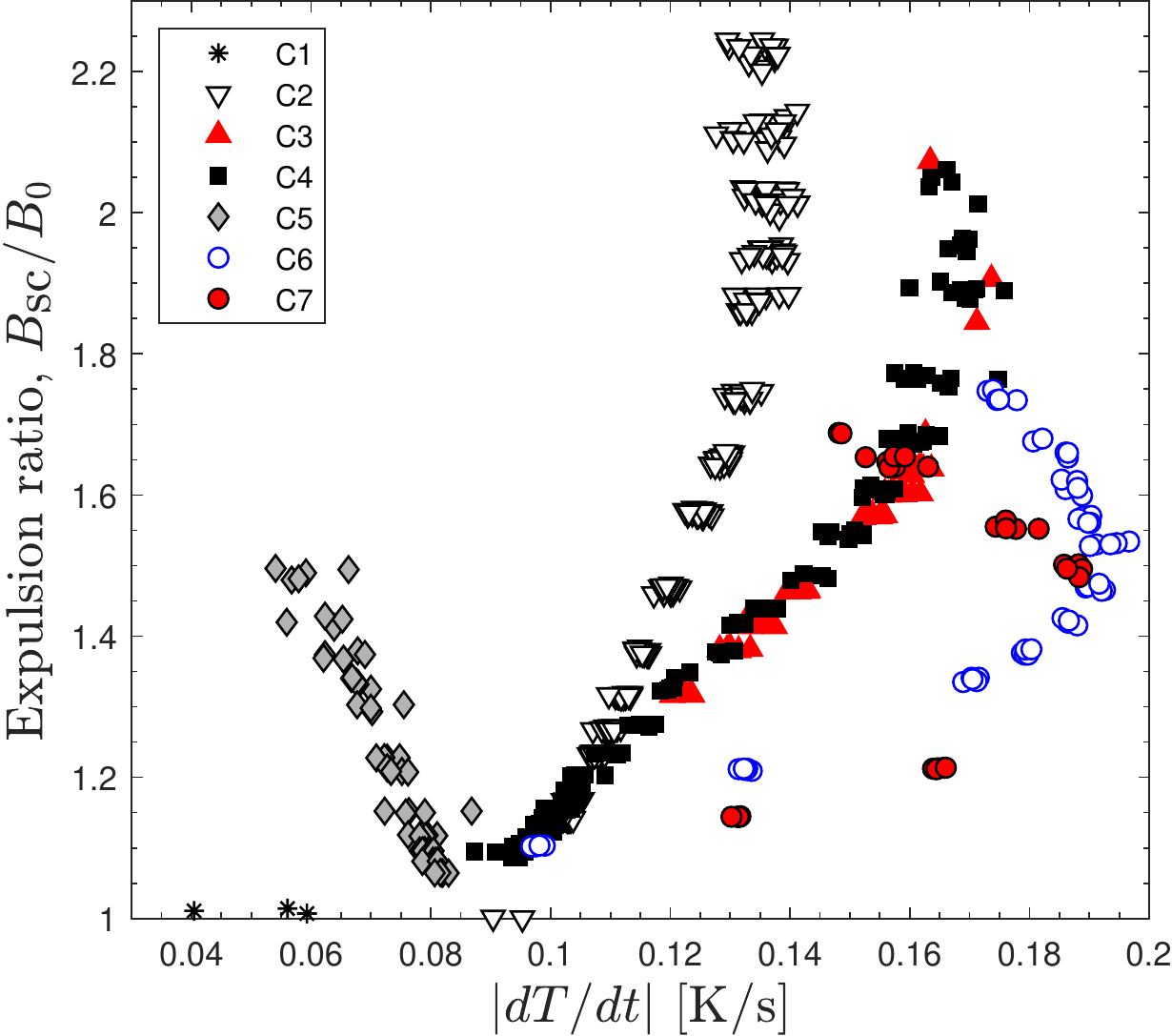}
	\caption{\label{EF_vs_cooling_rate} Expulsion ratio as a function of the cooling rate.}
	\includegraphics[scale=0.58]{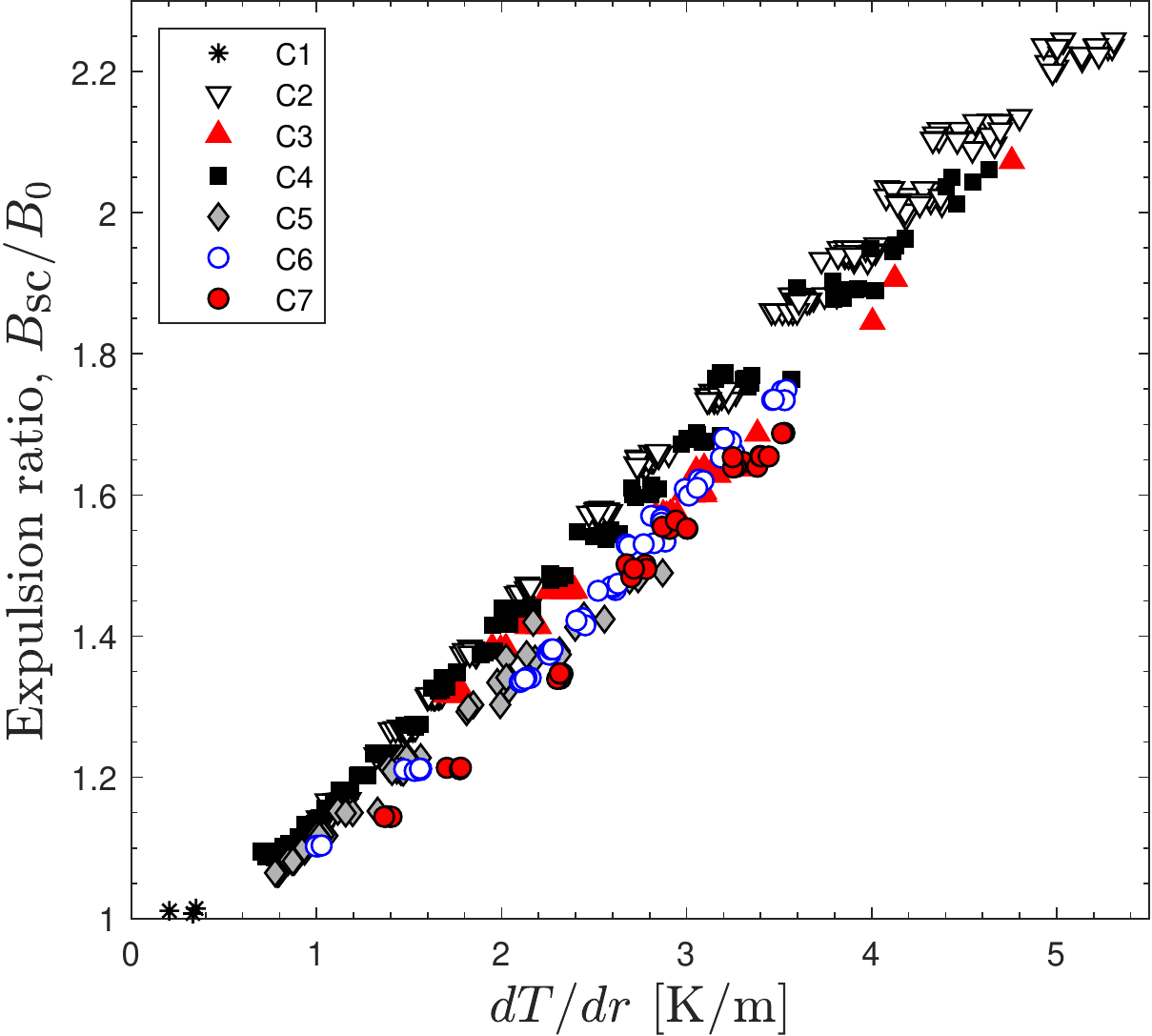}
	\caption{\label{EF_vs_grad} Expulsion ratio as a function of the spatial temperature gradient.}
\end{figure}
A total of seven measurement campaigns were performed (indexed as C1 to C7 and summarized in table \ref{results_table}). In each, a set of expulsions was made at varying $\nabla T$ where a total of 669 flux expulsions were obtained. The dependence of the ER as a function of cooling rate and the inverse front propagation speed is shown in figures \ref{EF_vs_front_speed} and \ref{EF_vs_cooling_rate}. The key result showing  ER as a function of the gradient is obtained via equation \ref{gradT} and is reported in figure \ref{EF_vs_grad}. 

\begin{table}
	\caption{\label{results_table}Measurement campaigns summarized.}
	\begin{ruledtabular}
		\begin{tabular}{cccccc}
			Campaign&  Expulsions & $\textrm{max}\{\nabla T\}$ [K/m]& $\textrm{max} \{  \frac{B_{sc}}{B_0} \}$ & Slope \footnote{In units of relative increase of flux density per K/m, calculated from a linear fit of the data in figure \ref{EF_vs_grad}.} \\
			\hline
			C1 & 3 &   0.4 & 1.01 & -- \\
			C2 & 300 & 5.3 & 2.24 & 0.27\\
			C3 & 75 &  4.8 & 2.07 & 0.23\\
			C4 & 144 & 4.8 & 2.06 & 0.26\\
			C5 & 57 &  2.9 & 1.5 &  0.22\\
			C6 & 55 &  3.5 & 1.75 & 0.26\\
			C7 & 35 &  3.5 & 1.69 & 0.27\\
		\end{tabular}
	\end{ruledtabular}
\end{table}

The details at which the measurements were performed are as follows: First, in C1 a very low gradient of 0.4 K/m was reached for which the obtained ER of $\approx$ 1 indicated nearly all flux was trapped in the sample. For C2 the thermal contact between the sample and the cold edge was improved by adding an indium gasket. This allowed to reach a maximum gradient of $\approx$ 5.3 K/m and ER of 2.24,  somewhat close to the value of 2 predicted by the simulations and meant sufficient part of the flux was expelled. To test reproducibility, for C3 the sample was remounted without replacing the indium gasket, and then for C4 the measurement was repeated after thermal cycling to room temperature but without  disassembly of the setup. In these cases we consistently reached slightly lower gradient of $\approx$ 4.8 K/m for which ER of $\approx$ 2 was obtained. Next, to investigate the role of the thermal contact resistance, in test C5 the force at the spring-loaded contact of the cold edge was nearly tripled. In this case a much lower maximum gradient of 2.9 K/m could be reached and a ER of $\approx$ 1.5 accordingly. Lastly, a more practical version of the indium gasket was developed that can be kept fixed in place allowing much easier installation. In C6 and C7 this new design was tested by remounting the sample and remeasuring. Around 3.5 K/m was reached with ER of 1.75 and 1.7 in the two campaigns correspondingly. 

Based on these sets of measurements,  the following observations and conculsions can be made:
\begin{enumerate}
	\item For each heating pulse length, expulsion was repeated multiple times (20 times in the majority of cases) which is the reason for the clustering of the points seen for the separate campaigns. Within a given measurement campaign the thermodynamic conditions at cool down are repeatable and can be thoroughly controlled solely by varying the heating pulse duration. 
	\item  Conditions from  campaign to campaign vary significantly, and this is seen clearly  in figures \ref{EF_vs_front_speed} and \ref{EF_vs_cooling_rate}, with each campaign differing in terms  of the  maximum thermal gradient that can be reached.  This is mainly attributed to the thermal contact near the cold edge: at temperatures near ${T}_{\textrm{c}}$ its thermal resistance is sensitive to the applied force, the amount of plastic deformation of the indium gasket and possibly to the surface finish of the sample \cite{Dhuley_2019} (e.g the result from C5 emphasizes the effect of the stiffer springs used). Since no special care was taken to control the thermal contact during the (done by hand) assembly nor was the surface specially treated, this variation in the thermodynamic behavior is unsurprising.
	\item When the increase of flux density is considered as a function of $\nabla T$ (shown in figure \ref{EF_vs_grad}) we see that the results from all campaigns converge in a consistent trend that quantifies how the expulsion improves as temperature gradient increases. The large number of expulsions repeatedly performed at varying thermodynamic conditions give confidence to conclude that spatial thermal gradient can be used to parameterise flux expulsion in this setup. This result demonstrates that our device can be used to characterize the intrinsic ability of the material to expel  magnetic flux. Our findings are fully in line with previously published studies performed on bare cavities \cite{Romanenko_2014,Posen_2016,Huang_Kubo_2016} and also underpin the conclusion that cooling at high rate is not uniquely related to high expulsion of trapped flux \cite{Huang_Kubo_2016}.
	\item  The small systematic offset seen for C6 and C7 in the final result is attributed to the changed design of the indium gasket. 
	\item The increased spread seen for the points of higher gradient comes from the evaluation of the cooling rate. In these cases the heating pulse is short and the sample is heated barely above $\textrm{T}_{\textrm{c}}$,  so that expulsion happens almost instantly after the temperature starts dropping. However, the thermal inertia of the sensor makes it difficult to follow the change of temperature potentially leading to a loss of resolution. 
	\item The maximum ER of $\approx$ 2.2 measured for $\nabla T$ of $ \approx$ 5.3 K/m is found greater than the value of \   $\approx$ 2 predicted from simulations. Moreover, no saturation is seen in figure \ref{EF_vs_grad} indicating complete flux expulsion was not reached. It is plausible that the low spatial resolution of the probe prevents precise measurement as the field varies significantly along the aperture axis. To overcome this uncertainty, we plan to use a search coil implemented as discussed in section \ref{concept}. In this way, the setup should in principle allow to directly quantify the absolute amount of flux expelled from the sample.
	\item The maximum spatial thermal gradient achieved may seem low and impractical compared to flux expulsion conditions for SRF cavity cool downs in vertical cryostats, which are up to an order of magnitude higher \cite{Romanenko_2014,Posen_2016,Huang_Kubo_2016}.  This would explain the lack of ER saturation, but does not prevent quantitative characterisation of material samples. The limit of the spatial thermal gradient for this proof-of-concept device is largely attributed to the cryogenic setup, with the cryocooler capacity limited to $\approx$ 1 W of cooling power. 
\end{enumerate}

\section{Summary}
\label{Summary}
This work introduced a new magnetometry measurement system to quantify magnetic flux expulsion from superconducting samples made of sheet material. By conduction cooling on a closed thermal topology, the flux expelled from the volume of a disc-shaped sample is collimated, permitting reproducible precision measurement of the expulsion efficiency.  Controlled pulse heating near the center of the sample is used to trigger the evolution of the superconducting front with the heating pulse length being the single knob to control the cool down dynamic. For simplicity of design and operation, this dedicated instrument has been  implemented using a commercial 1 W cryocooler.

A benchmark study performed with SRF cavity-grade RRR = 300 niobium, tested in seven measurement campaigns yielded more than 650 phase transitions in a range of cooling conditions:  cooling rate of (0.04 to 0.2) K/s; speed of the superconducting front of (2 to 18) cm/s; $\nabla T$ of (0.3 to 5.3) K/m. The main result of the study is a clear correlation  between the flux expulsion ratio and the spatial temperature gradient.  This result is both repeatable and reproducible, and  provides proof of concept that the flux expulsion lens can be used as a stand alone experiment to relate the amount of expelled flux to the cooling dynamics near $T_\textrm{c}$. Furthermore, these results show that high cooling rate is not uniquely related to strong expulsion of trapped flux, and clarifies the point that the spatial temperature gradient rather than the cooling rate has the key role to improve expulsion of magnetic flux. 

By developing a stand-alone setup to test material samples in a well-controlled thermal environment  a standardized measurement of flux expulsion can be achieved.  Our device allows for such measurements to be added to SRF cavity performance parameter matrix, both for raw sheet material and work-hardened samples.  In addition, as the design suppresses extrinsic thermoelectric effects,  the  setup can potentially address thermoelectric effects intrinsic to superconducting thin film samples such as niobium on a copper substrate. Due to its simplicity, the flux lens may also find its application in other areas where expulsion of flux is critical such as superconducting circuits or magnetic shielding based on superconductors. 

\begin{acknowledgments}
	This project has received funding from the European Union’s Horizon 2020 Research and Innovation program under Grant Agreement No 730871. We would also like to thank Frank Gerigk (CERN), Walter Venturini Delsolaro (CERN) and Akira Miyazaki (Uppsala University) for the numerous fruitful discussions, as well as to Tommi Mikkola (CERN) for his expertise in the mechanical design. 
\end{acknowledgments}

\bibliography{v3}

\end{document}